\documentstyle[sprocl,epsfig]{article}

\def\be{\begin{equation}}
\def\ee{\end{equation}}
\def\bea{\begin{eqnarray}}
\def\eea{\end{eqnarray}}

\begin{document}

\title{
  CP and T violation in long baseline experiments with low energy
neutrino
}

\author{Joe Sato}

\address{Research Center for Higher Education, Kyushu University,\\
Ropponmatsu, Chuo-ku, Fukuoka, 810-8560\\
E-mail: joe@rc.kyushu-u.ac.jp}


\maketitle
\abstracts{
   Stimulated by the idea of PRISM, a very high intensity muon ring
   with rather low energy, we consider possibilities of observing
   CP-violation effects in neutrino oscillation experiments.
   Destructive sum of matter effect and
   CP-violation effect can be avoided with use of initial $\nu_{\rm
   e}$ beam.  We finally show that the experiment with (a few)
   $\times$ 100 MeV of neutrino energy and (a few) $\times$ 100 km of
   baseline length, which is considered in this paper, is particularly
   suitable for a search of CP violation in view of 3-generation
   nature of CP violation.
}

\section{Introduction}

Many experiments and observations have shown evidences for neutrino
oscillation one after another.  The solar neutrino deficit has long
been observed\cite{Ga1,Ga2,Kam,Cl,SolSK}.  The atmospheric neutrino
anomaly has been found\cite{AtmKam,IMB,SOUDAN2,MACRO} and recently
almost confirmed by SuperKamiokande\cite{AtmSK}.  There is also
another suggestion given by LSND\cite{LSND}.  All of them can be
understood by neutrino oscillation and hence indicates that neutrinos
are massive and there is a mixing in lepton
sector\cite{FukugitaYanagida}.

Since there is a mixing in lepton sector, it is quite natural to
imagine that there occurs CP violation in lepton sector.  Several
physicists have considered whether we may see CP-violation effect in
lepton sector through long baseline neutrino oscillation experiments%
\cite{CPCite}.

The use of neutrinos from muon beam has great advantages compared with
those from pion beam.  Neutrinos from $\mu^+$($\mu^-$) beam consist of
pure $\nu_{\rm e}$ and $\bar\nu_\mu$ ($\bar\nu_{\rm e}$ and $\nu_\mu$)
and will contain no contamination of other kinds of neutrinos.  Also
their energy distribution will be determined very well.  In addition
we can test T violation in long baseline experiments by using
(anti-)electron neutrino\cite{ArafuneJoe,AKS}.

What energy range is suitable for observing CP violation?
 Since CP-violation effect arise as three(or more)-generation
phenomena\cite{KM,Cabibbo,BWP}, we should make an experiment
with ``not too high'' and ``not too low'' energy to see ``3-generation''.
In an oscillation experiment, there are two energy scales,
\begin{eqnarray}
E\ \sim\ 
\left\{
\begin{array}{l}
\ \delta m^2_{31} L \\
\ \delta m^2_{21} L
\end{array}
\right. .
\label{energyRange}
\end{eqnarray}
Then the above energy range is expected to be suitable for
a neutrino oscillation experiment to see CP violation in lepton
sector.

More to say, to avoid matter effect which gives a fake CP violation,
the lower energy is expected to be more preferable. Though
unfortunately neutrinos in neutrino factory seem to have 
very high energy\cite{Geer}, very luckily we will have very intense muon
source with rather low energy, PRISM\cite{PRISM}.  It will located at
Tokai, Ibaraki Prefecture, about 50 km from KEK. Since the muons will
have energy less than 1 GeV, we can expect that we will have very
intense neutrino beam with energy less than 500 MeV. With baseline
length of several hundreds km, it will be very
suitable to explore CP violation in lepton sector with neutrino
oscillation experiments.  With such a low energy beam, we will be able
to detect neutrinos experimentally with good energy resolution. 
Stimulated by the possibility that we will have a low energy neutrino
source with very high intensity, we consider here how large
CP-violation effect we will see with such neutrino beam.

In this paper we will consider three active neutrinos without any
sterile one by attributing the solar neutrino deficit and atmospheric
neutrino anomaly to the neutrino oscillation\cite{KSPrism}.
We will use the following notation for the mixing matrix $U$,
\begin{eqnarray}
& &
U
=
{\rm e}^{{\rm i} \psi \lambda_{7}} \Gamma {\rm e}^{{\rm i} 
\phi \lambda_{5}} {\rm e}^{{\rm i} \omega \lambda_{2}} \nonumber 
\\
&=&
\left(
\begin{array}{ccc}
  1 & 0 & 0  \\
  0 & c_{\psi} & s_{\psi} \\
  0 & -s_{\psi} & c_{\psi}
\end{array}
\right)
\left(
\begin{array}{ccc}
  1 & 0 & 0  \\
  0 & 1 & 0  \\
  0 & 0 & {\rm e}^{{\rm i} \delta}
\end{array}
\right)
\left(
\begin{array}{ccc}
  c_{\phi} & 0 &  s_{\phi} \\
  0 & 1 & 0  \\
  -s_{\phi} & 0 & c_{\phi}
\end{array}
\right)
\left(
\begin{array}{ccc}
  c_{\omega} & s_{\omega} & 0 \\
  -s_{\omega} & c_{\omega} & 0  \\
  0 & 0 & 1
\end{array}
\right)
\nonumber \\
&=&
\left(
\begin{array}{ccc}
   c_{\phi} c_{\omega} &
   c_{\phi} s_{\omega} &
   s_{\phi}
  \\
   -c_{\psi} s_{\omega}
   -s_{\psi} s_{\phi} c_{\omega} {\rm e}^{{\rm i} \delta} &
   c_{\psi} c_{\omega}
   -s_{\psi} s_{\phi} s_{\omega} {\rm e}^{{\rm i} \delta} &
   s_{\psi} c_{\phi} {\rm e}^{{\rm i} \delta}
  \\
   s_{\psi} s_{\omega}
   -c_{\psi} s_{\phi} c_{\omega} {\rm e}^{{\rm i} \delta} &
   -s_{\psi} c_{\omega}
   -c_{\psi} s_{\phi} s_{\omega} {\rm e}^{{\rm i} \delta} &
   c_{\psi} c_{\phi} {\rm e}^{{\rm i} \delta}
\end{array}
\right),
\label{UPar2}
\end{eqnarray}
where $c_{\psi} = \cos \psi, s_{\phi} = \sin \phi$, etc,
and matter effect\cite{Wolf,MS} $a$,
\begin{equation}
 a \equiv 2 \sqrt{2} G_{\rm F} n_{\rm e} E \nonumber \\
   = 7.56 \times 10^{-5} {\rm eV^{2}} \cdot
       \left( \frac{\rho}{\rm g\,cm^{-3}} \right)
       \left( \frac{E}{\rm GeV} \right).
 \label{aDef}
\end{equation}

We will assume above energy range $\sim$ several hundreds MeV
and hence from (\ref{energyRange}) with baseline length
$\sim$ several hundreds km. With such an experimental setting
the oscillation probabilities are calculated, e.g., as\cite{AKS}
\begin{eqnarray}
& &
 P(\nu_{\mu} \rightarrow \nu_{\rm e}; E, L)
=
 4 \sin^2 \frac{\delta m^2_{31} L}{4 E}
 c_{\phi}^2 s_{\phi}^2 s_{\psi}^2
 \left\{
  1 + \frac{a}{\delta m^2_{31}} \cdot 2 (1 - 2 s_{\phi}^2)
 \right\}
 \nonumber \\
&+&
 2 \frac{\delta m^2_{31} L}{2 E} \sin \frac{\delta m^2_{31} L}{2 E}
 c_{\phi}^2 s_{\phi} s_{\psi}
 \left\{
  - \frac{a}{\delta m^2_{31}} s_{\phi} s_{\psi} (1 - 2 s_{\phi}^2)
  +
 \frac{\delta m^2_{21}}{\delta m^2_{31}} s_{\omega}
    (-s_{\phi} s_{\psi} s_{\omega} + c_{\delta} c_{\psi} c_{\omega})
 \right\}
 \nonumber \\
&-&
 4 \frac{\delta m^2_{21} L}{2 E} \sin^2 \frac{\delta m^2_{31} L}{4 E}
 s_{\delta} c_{\phi}^2 s_{\phi} c_{\psi} s_{\psi} c_{\omega}
 s_{\omega}.
 \label{eq:AKSmu2e}
\end{eqnarray}

\section{CP violation search in long baseline experiments}

\subsection{Magnitude of CP violation and matter effect}

The available neutrinos as an initial beam are $\nu_{\mu}$ and
$\bar\nu_{\mu}$ in the current long baseline
experiments\cite{K2K,Ferm}.  The ``CP violation'' gives the nonzero
difference of the oscillation probabilities between, e.g.,
$P(\nu_{\mu} \rightarrow \nu_{\rm e})$ and $ P(\bar\nu_{\mu}
\rightarrow \bar\nu_{\rm e})$\cite{AKS}.  This gives
\begin{eqnarray}
 P(\nu_{\mu} \rightarrow \nu_{\rm e}; L)
-
 P(\bar\nu_{\mu} \rightarrow \bar\nu_{\rm e}; L)
&=&
 16 \frac{a}{\delta m^2_{31}} \sin^2 \frac{\delta m^2_{31} L}{4 E}
 c_{\phi}^2 s_{\phi}^2 s_{\psi}^2 (1 - 2 s_{\phi}^2)
 \nonumber \\
&-&
 4 \frac{a L}{2 E} \sin \frac{\delta m^2_{31} L}{2 E}
 c_{\phi}^2 s_{\phi}^2 s_{\psi}^2 (1 - 2 s_{\phi}^2)
 \nonumber \\
&-&
 8 \frac{\delta m^2_{21} L}{2 E}
 \sin^2 \frac{\delta m^2_{31} L}{4 E}
 s_{\delta} c_{\phi}^2 s_{\phi} c_{\psi} s_{\psi} c_{\omega}
 s_{\omega}.
 \label{eq:CP}
\end{eqnarray}
The difference of these two, however, also includes matter effect, or
the fake CP violation, proportional to $a$.  We must somehow
distinguish these two to conclude the existence of CP violation as
discussed in ref.\cite{AKS}.

On the other hand, a muon ring enables to extract $\nu_{\rm e}$ and
$\bar\nu_{\rm e}$ beam.  It enables direct measurement of
pure CP violation through ``T violation'',
e.g., $P(\nu_{\mu} \rightarrow
\nu_{\rm e}) - P(\nu_{\rm e} \rightarrow \nu_{\mu})$ as
\begin{equation}
 P(\nu_{\mu} \rightarrow \nu_{\rm e}) -
 P(\nu_{\rm e} \rightarrow \nu_{\mu})
=
 - 8 \frac{\delta m^2_{21} L}{2 E}
 \sin^2 \frac{\delta m^2_{31} L}{4 E}
 s_{\delta} c_{\phi}^2 s_{\phi} c_{\psi} s_{\psi} c_{\omega}
 s_{\omega}.
 \label{eq:T}
\end{equation}
Note that this difference gives pure CP violation.

By measuring ``CPT violation'', e.g.
the difference between $P(\nu_{\mu} \rightarrow 
\nu_{\rm e})$ and $P(\bar\nu_{\rm e} \rightarrow \bar\nu_{\mu})$,
we can check the matter effect.
\begin{eqnarray}
 P(\nu_{\mu} \rightarrow \nu_{\rm e}; L)
-
 P(\bar\nu_{\rm e} \rightarrow \bar\nu_{\mu}; L)
&=&
 16 \frac{a}{\delta m^2_{31}} \sin^2 \frac{\delta m^2_{31} L}{4 E}
 c_{\phi}^2 s_{\phi}^2 s_{\psi}^2 (1 - 2 s_{\phi}^2)
 \nonumber \\
&-&
 4 \frac{a L}{2 E} \sin \frac{\delta m^2_{31} L}{2 E}
 c_{\phi}^2 s_{\phi}^2 s_{\psi}^2 (1 - 2 s_{\phi}^2)
 \nonumber \\
 \label{eq:CPT}
\end{eqnarray}
%


\begin{figure}
 \unitlength=1cm
 \begin{picture}(15,18)
  \unitlength=1mm
  \centerline{
   \epsfysize=18cm
   \epsfbox{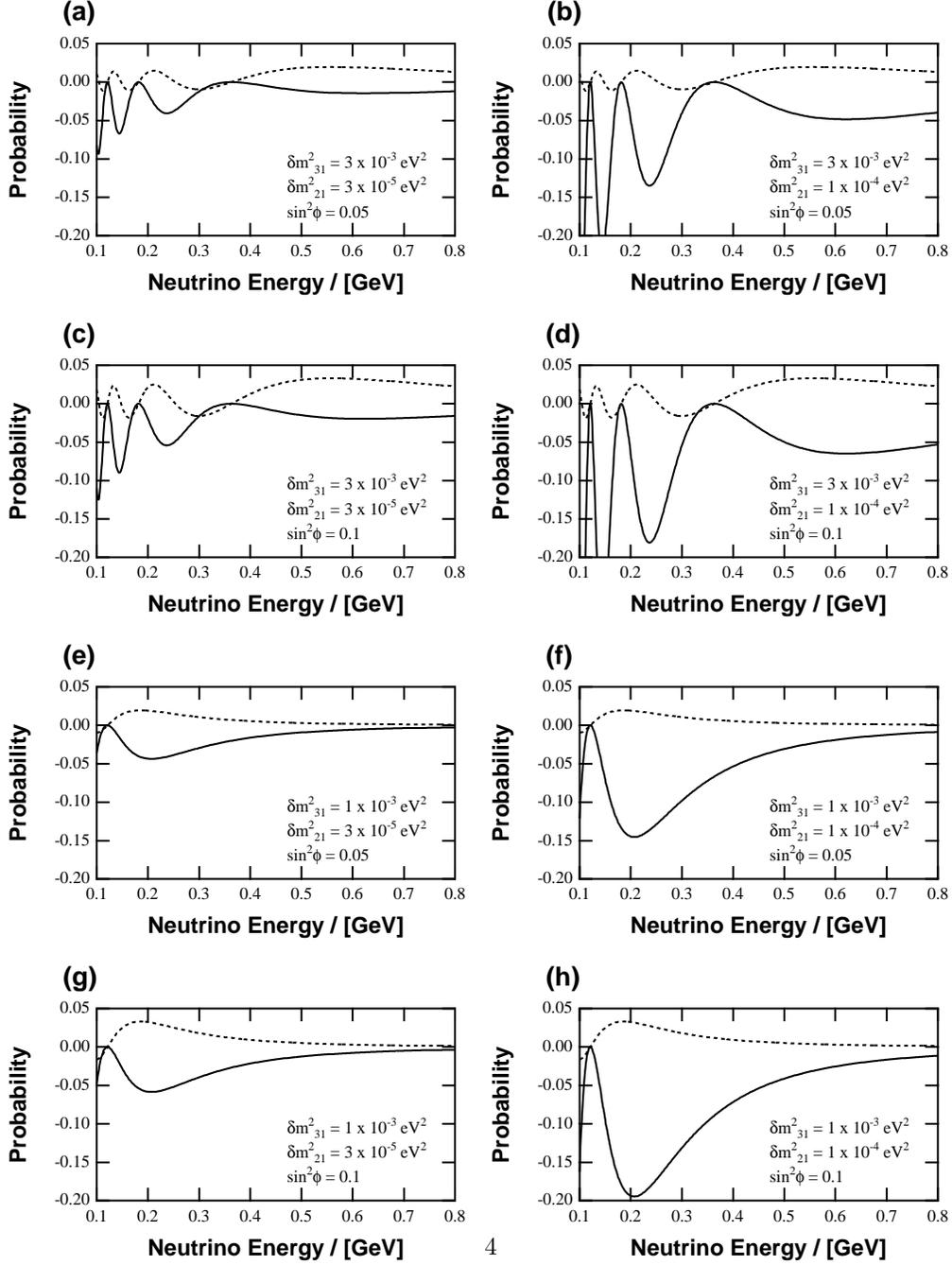}
  }
 \end{picture}
\caption{ Graphs of $P(\nu_{\mu} \rightarrow \nu_{\rm e}) - P(\nu_{\rm
e} \rightarrow \nu_{\mu})$ (solid lines; pure CP-violation effects) and
$P(\nu_{\mu} \rightarrow \nu_{\rm e}) - P(\bar\nu_{\rm e} \rightarrow
\bar\nu_{\mu})$ (dashed lines; matter effects) as functions of neutrino
energy.  Parameters not shown in the graphs are taken 
$\sin^{2} \omega = 1/2, \sin^{2} \psi = 1/2, \sin \delta
= 1; \rho = {\rm g/cm^{3}}$ and $L = 300 {\rm km}$.} \label{fig2}
\end{figure}


We present in Fig.\ref{fig2} ``T-violation'' part (\ref{eq:T}) and
``CPT-violation'' part (\ref{eq:CPT}) for some parameters allowed by
the present experiments\cite{Fogli} with $\sin^{2}
\omega = 1/2$, $\sin^{2} \psi = 1/2$, $\sin \delta = 1$ fixed%
\footnote{Although the Chooz reactor
experiment have almost excluded $\sin^2\phi=0.1$\cite{chooz}, there
remains still small chance to take this value.}.  The
matter density is also fixed to the constant value\cite{KS}
$\rho = 2.5 {\rm
g/cm^{3}}$.  Other parameters are taken as $\delta m^{2}_{31}
= 3 \times 10^{-3} {\rm eV}^{2}$ and $1 \times 10^{-3} {\rm eV}^{2}$,
$\delta m^{2}_{21} = 1 \times 10^{-4} {\rm eV}^{2}$ and $3 \times
10^{-5} {\rm eV}^{2}$.

``T-violation'' effect is proportional to $\delta m^{2}_{21} / \delta
m^{2}_{31}$ and, for $\phi \ll 1$, also to $\sin \phi$ as seen in
eq.(\ref{eq:T}) and Fig.\ref{fig2}.  Recalling that the energy of neutrino 
beam
is of several hundreds MeV, we see in Fig.\ref{fig2} that the
``T-violation'' effect amounts to at least about 5\%, hopefully
10$\sim$20\%.  This result gives hope to detect the pure leptonic CP
violation directly with the neutrino oscillation experiments.

The ``T violation'' is, however, less than 10\% in the case that
$\delta m^{2}_{21}$ is as small as $3 \times 10^{-5} {\rm eV}^{2}$
(see the left four graphs of Fig.\ref{fig2}).  In this case matter
effect is as large in magnitude as ``T violation'' and has an opposite
sign for $\sin\delta>0$ as seen in Fig.\ref{fig2}.  In such a case the
sum of the two, eq.(\ref{eq:CP}), is destructive and has even more
smaller magnitude than ``T violation'', thus the experiments will be
more difficult.  Thanks to $\nu_{\rm e}$ and $\bar\nu_{\rm e}$
available from low energy muon source, one can measure ``T
violation''.  This makes the measurement much easier.

\subsection{Estimation of statistical error in CP-violation searches}

Here we state that the energy range considered here is probably best
in view of statistical errors in order to observe CP violation effect. 
To this end let us estimate how $\delta P / \Delta P$ scales with $E$
and $L$, where $\delta P$ be statistical error of transition
probabilities such as $P(\nu_{\rm e} \rightarrow \nu_{\mu})$ and
$\Delta P = P(\nu_{\rm e} \rightarrow \nu_{\mu}) - P(\nu_{\mu}
\rightarrow \nu_{\rm e})$.  We denote in this section the transition
probabilities $P(\nu_{\alpha} \rightarrow \nu_{\beta}) (\alpha \ne
\beta)$ simply by $P$.  Suppose that $n$ neutrinos out of $N$ detected
neutrinos has changed its flavor.  With a number of decaying muons
fixed, the number of detected neutrinos $N$ are roughly proportional
to $E^3$, and hence $N \sim E^{3} L^{-2}$.  We estimate $\delta P$ as

\begin{eqnarray}
    \delta P &=&
    \delta \left( \frac{n}{N} \right)
    \nonumber \\
    &=&
    \frac{|N \delta n| + |n \delta N|}{N^{2}}
    \nonumber \\
    &=&
    \frac{|N \sqrt{NP}| + |NP \sqrt{N}|}{N^{2}}
    \nonumber \\
    &=&
    \frac{\sqrt{P} + P}{\sqrt{N}},
    \label{eq:deltaP}
\end{eqnarray}
where we used $\delta n = \sqrt{n}$, $\delta N = \sqrt{N}$ and $n = N
P$.  From eqs.(\ref{eq:AKSmu2e}), (\ref{eq:T}) and (\ref{eq:deltaP}),
we can estimate how $\delta P / \Delta P$ scales for $E$ with $L$
fixed.  We summarize the results in Table \ref{table:Escale}.  There
we see that $\delta P / \Delta P$ reaches minimum at the region $E
\sim \delta m^{2}_{31} L$. Note that
indeed the 3-generation nature (\ref{energyRange})
 is satisfied here. With such a concrete example,
we can see how important the 3-genaration nature is.
\begin{table}
    \begin{center}
\begin{tabular}{|c||c|c|c|c|c|}
    \hline
    $E$ &  & $\delta m^{2}_{21} L$ & & $\delta m^{2}_{31} L $ & \\ 
    \hline \hline
    $P$ & ``const.'' &  & $1/E$ or ``const.'' & & $1/E^{2}$ \\
    \hline
    $\delta P$ & $1/E^{1.5}$ & & $1/E^{1.5} \sim 1/E^{2}$ & & $1/E^{2.5}$ \\
    \hline
    $\Delta P$ & ``const'' & & $1/E$ & & $1/E^{3}$ \\
    \hline \hline
    $\delta P / \Delta P$ & $1/E^{1.5}$ &  & $1/E^{0.5} \sim 1/E$ & & $E^{0.
5}$ 
    \\
    \hline
     & $\searrow$ & & $\searrow$ & minimum & $\nearrow$ \\
    \hline
\end{tabular}
\caption{The $E$-dependence of oscillation envelopes of some quantities
with $L$ fixed.  Here ``const.''  means that the oscillation envelope
of the quantity is independent of $E$.  $\delta P / \Delta P$ reaches
minimum at the region $E \sim \delta m^{2}_{31} L$.}
\label{table:Escale}
\end{center}
\end{table}

By a similar consideration one can obtain how $\delta P / \Delta P$ 
scales for $L$ with $E$ fixed.  The result for this case is shown in 
Table \ref{table:Lscale}.  We can see there that we should keep not 
too large $L$ so that the error $\delta P / \Delta P$ should not get 
large.
\begin{table}
    \begin{center}
\begin{tabular}{|c||c|c|c|c|c|}
    \hline
    $L$ &  &  $E/\delta m^{2}_{31} $ & & $E/\delta m^{2}_{21} $ & \\ 
    \hline \hline
    $P$ & $L^{2}$ &  & $L$ or ``const.'' & & ``const.'' \\
    \hline
    $\delta P$ & $L^{3}$ & & $L^{1.5} \sim L$ & & $L$ \\
    \hline
    $\Delta P$ & $L^{3}$ & & $L$ & & ``const.'' \\
    \hline \hline
    $\delta P / \Delta P$ & ``const.'' &  & $L^{0.5} \sim$ ``const.'' & & 
    $L$
    \\
    \hline
     & $\rightarrow$ &  & $\nearrow$ & & $\nearrow$ \\
    \hline
\end{tabular}
\caption{The $L$-dependence of oscillation envelopes of some quantities
with $E$ fixed.}
\label{table:Lscale}
\end{center}
\end{table}

We need a few hundreds MeV of neutrino energy to reach the threshold
energy of muon production reaction ${\rm N} + \nu_{\mu} \rightarrow
{\rm N} + \mu$, where N is nucleon.  We have also seen in
Table \ref{table:Escale} that the error comes to minimum at the region
$E \sim \delta m^{2}_{31} L$.  Considering these results, we conclude
that $E \sim$ (a few) $\times$ 100 MeV and $L \sim $ (a few) $\times$
100 km, which we have just considered in this paper, is the best
configuration to search CP violation in view of statistical error.

\section{Summary and conclusion}
We considered how large CP/T violation effects can be observed making
use of low-energy neutrino beam, inspired by PRISM. More than 10\%,
hopefully 20\% of the pure CP-violation effects may be observed within
the allowed region of present experiments.

We have also seen that in some case the pure CP-violation effects are
as small as the matter effect but have opposite sign.  In such a case
the ``CP violation'' gets smaller through the destructive sum of the
pure CP-violation effect and matter effect.  We pointed out that we
can avoid this difficulty by observing ``T-violation'' effect using
initial $\nu_{\rm e}$ beam.

We finally discussed that the configuration we have considered here,
$E \sim$ (a few) $\times$ 100 MeV and $L \sim$ (a few) $\times$ 100 km
is best to search lepton CP violation in terms of statistical error. 
With such consideration we also found how important the 3-generation
nature (\ref{energyRange}) is.
It is thus worth making an effort to develop leptonic CP violation
search using neutrinos from low energy muons.

\section*{Acknowledgments}
The author thanks M. Koike for useful discussions.
This research was supported in
part by a Grant-in-Aid for Scientific Research of the Ministry of
Education, Science and Culture, \#12047221, \#12740157.


\section*{References}
\begin{thebibliography}{99}
\bibitem{Ga1} GALLEX Collaboration, W.~Hampel {\it et al.}, Phys.
    Lett. B {\bf 447}, 127 (1999) .
    
\bibitem{Ga2} SAGE Collaboration, J.~N.~Abdurashitov {\it et al.},
    astro-ph/9907113.
    
\bibitem{Kam} Kamiokande Collaboration, Y.~Suzuki, Nucl. Phys. B (Proc.
    Suppl.) {\bf 38}, 54 (1995).
    
\bibitem{Cl} Homestake Collaboration, B.~T.~Cleveland {\it et al.},
  Astrophys.  J. {\bf 496}, 505 (1998).


\bibitem{SolSK} Super-Kamiokande Collaboration, Y. Fukuda {\it et
  al.}, Phys. Rev. Lett. {\bf 82},1810 (1999), {\it ibid.} {\bf 82},
  2430 (1999).

\bibitem{AtmKam} Kamiokande Collaboration, K.~S.~Hirata {\it et al.},
  Phys. Lett.  {\bf B205},416 (1988); {\it ibid.}  {\bf B280},146 (1992);
  Y.~Fukuda {\it et al.}, Phys. Lett.  {\bf B335}, 237 (1994).

\bibitem{IMB} IMB Collaboration, D.~Casper {\it et al.},
  Phys. Rev. Lett. {\bf 66}, 2561 (1991);\\ R.~Becker-Szendy {\it et
  al.}, Phys. Rev.  {\bf D46}, 3720  (1992).

\bibitem{SOUDAN2} SOUDAN2 Collaboration, T.~Kafka, Nucl. Phys. B
  (Proc.  Suppl.) {\bf 35}, 427 (1994); M.~C.~Goodman, {\it ibid.} {\bf
  38}, 337 (1995); W.~W.~M.~Allison {\it et al.}, Phys. Lett. {\bf
  B391}, 491 (1997).

\bibitem{MACRO} MACRO Collaboration, M. Ambrosio {\it et al.},
Phys. Lett. {\bf B434}, 451 (1998). 

\bibitem{AtmSK} Super-Kamiokande Collaboration, Y. Fukuda {\it et
  al.}, Phys. Rev. Lett. {\bf 81}, 1562 (1998), Phys. Lett.  {\bf B433},
  9 (1998), Phys. Lett.  {\bf B436}, 33 (1998), Phys. Rev. Lett. {\bf
  82},2644 (1999).

\bibitem{LSND} LSND Collaboration, C.Athanassopoulos {\it et al.},
 Phys. Rev. Lett {\bf 77}, 3082 (1996); {\it ibid} {\bf 81},
1774 (1998).

\bibitem{FukugitaYanagida} For a review, M. Fukugita and T. Yanagida,
    in {\it Physics and Astrophysics of Neutrinos}, edited by M.
    Fukugita and A. Suzuki (Springer-Verlag, Tokyo, 1994).

\bibitem{CPCite} see e.g. references in ref\cite{KSPrism}.

\bibitem{ArafuneJoe} J. Arafune and J. Sato, Phys. Rev.  {\bf D55},
    1653 (1997).

\bibitem{AKS} J. Arafune, M. Koike and J. Sato, 
    Phys. Rev. {\bf D56}, 3093 (1997).

\bibitem{KM} M. Kobayashi and T. Maskawa,
       Prog. Theor. Phys. 49, 652 (1973).

\bibitem{Cabibbo} N. Cabibbo, Phys. Lett. {\bf B72}, 333 (1978).

\bibitem{BWP} V. Barger, K. Whisnant and R. J. N. Phillips, Phys. Rev. 
Lett. {\bf 45}, 2084 (1980).

\bibitem{Geer} S. Geer, Phys. Rev. {\bf D57}, 6989 (1998), erratum
    {\it ibid.} {\bf D59}, 039903 (1999).

\bibitem{PRISM} Y. Kuno and Y. Mori, Talk at the ICFA/ECFA Workshop
    ``Neutrino Factories based on Muon Storage Ring'', July 1999;
    Y. Kuno, in {\it Proceedings of the Workshop on High Intensity
    Secondary Beam with Phase Rotation}.

\bibitem{KSPrism} M. Koike and J. Sato, Phys. Rev. {\bf D61} 073012 (2000).

\bibitem{Wolf} L. Wolfenstein, Phys. Rev. {\bf D17}, 2369 (1978).

\bibitem{MS} S. P. Mikheev and A. Yu. Smirnov,
    Sov. J. Nucl. Phys. {\bf 42}, 913 (1985).

\bibitem{K2K} K. Nishikawa, INS-Rep-924 (1992).
 
\bibitem{Ferm} S. Parke, Fermilab-Conf-93/056-T (1993),
    hep-ph/9304271.



\bibitem{Fogli} G. L. Fogli, E. Lisi, D. Marrone and G.  Scioscia,
    Phys. Rev. D {\bf 59}, 033001 (1999).

\bibitem{chooz} M. Apollonio {\it et al.}, Phys. Lett. {\bf B420},
 397 (1998); hep-ex/9907037.



\bibitem{KS} M. Koike and J.Sato, Mod. Phys. Lett. {\bf A14},1297 (1999).

\end{thebibliography}
\end{document}